7/21/20

# Using math in physics:
# 1. *Dimensional analysis*

*Edward F. Redish,*
University of Maryland - emeritus, College Park, MD

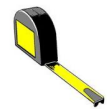 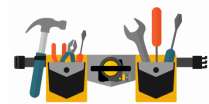

Making meaning with math in physics requires blending physical conceptual knowledge with mathematical symbology. Students in introductory physics classes often struggle with this, but it is an essential component of learning how to think with math. Teaching *dimensional analysis* (DA) — figuring out what measurements were combined to create a symbolic quantity — is a valuable first step in helping them learn to appreciate this difference. In this paper I discuss some of the issues associated with learning dimensional analysis and show some ways we can modify our instruction to help. This paper is one of a series on how to help students develop the scientific thinking skills required for learning to use math in science.[1]

We often treat DA as if it's just unit checks and only a way to find calculational mistakes. But dimensionality[2] plays a more fundamental conceptual role. DA is one of the basic "e-games" (knowledge building strategies or "epistemic games")[3] that can help students learn to blend physical concepts with mathematical representations. For our pre-medical students, the AAMC, the group responsible for developing the MCAT, has identified dimensional analysis as one of the primary learning objectives in developing quantitative numeracy.[4]

The icon I use for the DA e-game is a measuring tape shown at the top. Every time I use DA in class that icon appears on the slide. Every time it's used in our text (a free web-based wiki[5]) the icon appears. This provides a visual marker to remind students how valuable (and common) this strategy is.

## Measurements are not numbers

The symbols we use in physics are typically not "numbers" but represent physical quantities that can be assigned a number in a variety of ways, depending on choices we get to make. Whenever we have a free choice (as of a unit), we can introduce a "measurement dimension" or *dimensionality* that specifies what kinds of measurement we used to generate our results. Did we use a tape measure marked in inches or one in centimeters? A clock measured in seconds or hours? Each of these is a choice and assigning a number to the measurement implies a choice of standard or *unit* of length. When we have such a choice, we have to be careful to specify the unit we've chosen so anyone we're giving our number to knows what it means physically.[6]

When we write an equation containing measurements, the statement that two things are equal means that they match physically, not that they have the same number. They will have the same numerical value only if they are expressed in the same units. As a result, dimensioned equations can look peculiar if you're only thinking about the math of pure numbers.

The equation 1 inch = 2.54 cm is a legitimate equation in physics because both sides represent the same physical length. The equation $x = t$ with $x$ = 3 cm and $t$ = 3 sec is not a legitimate equation even though the numbers match. If we choose different units (as we are free to), the numbers no longer match. A length of 3 cm is also equal to a length of 1.18 inches. A time of 3 seconds can also be written at a time of 0.05 minutes. While $3 = 3$, $1.18 \neq 0.05$. If we are thinking about $x$ as a distance and $t$ as a time, it's pretty clear that distance and time are two distinct things.

The basic principle is:

> *We can only equate (or add) quantities that are the same kind of thing (have the same dimensionality) — that change the same way when we change our choice of unit .*

In a physically valid equation, the units don't have to match and the numbers don't have to match, but the dimensionalities do. This hints that something fundamental is going on. We're actually hiding some sophisticated math. Our symbols are not numbers but quantities that transform in particular ways when we change the arbitrary choice we are free to make. For example, if we change our standard unit by a factor of $\lambda$, then the number assigned to a quantity measured directly by our measurement scale will change by a factor of $1/\lambda$. If we change our scale from centimeters to meters (100 times bigger) the numbers we assign to lengths will get 100 times smaller.

If we want our equations to be physically meaningful (not just mathematically meaningful), both sides of the equation have to change in the same way when we choose a different scale. If we have a combination of different kinds of measurements, both sides of an equation (or terms we are adding) have to change in the same way when we change our choice of scale for any of the measurements. This means that when we combine measurements — a volume as a product of three length measurements, or a velocity as the ratio of a length measurement to a time measurement — we can only equate (or add) quantities of the same type: volumes with volumes, velocities with velocities.

Understanding how we use symbols to represent a measurement is the first step in blending a physical concept with a mathematical one. We are assigning a number to a symbol, but it's not a fixed number. What's fixed is a property of the physical object we are describing. This is a rather dramatic conceptual shift and one many students have trouble making.[7]

In my introductory physics class, I use five different kinds of measurements:

- A measurement made with a ruler (L, a length) 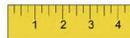
- A measurement made with a clock (T, a time)
- A measurement made with a scale (M, a mass)
- A measurement of electric strength (Q, a charge)
- A measurement made with a thermometer ($\Theta$, a temperature) 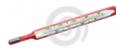

## Dimensionalities are important for a number of reasons

1. DA is a good first step in learning to blend physical and mathematical concepts.

DA is a good way to begin learning to think of symbols as having a physical correlate instead of being just a number.

2. DA is a good way to check that your equations do not contain an error.

Students often approach equations as something to memorize. Since recall is constructed and not always accurate,[8] students can "cross up" two equations and produce an incorrect or inappropriate one. Checking dimensionalities helps them disentangle memory errors.

3. DA is a good way to focus on functional dependence.

One of the most important ideas that comes out of DA is that it is not just the correlation in dependence that matters ("more A means more B, but more A means less C") but *how* it depends. Two powerful examples are surface vs volume scaling and the difference between viscous drag (proportional to v) and inertial drag (proportional to $v^2$). The idea of how quantities depend on each other — functional dependence and scaling — is so important that I define it as a separate e-game.[9]

4. DA is a good way to generate possible equations for new physical situations.

If we are teaching life-science students, it is important for us to cover such complex topics as fluids where the derivation of equations for viscous flow (Hagen-Poiseuille) or capillary action may be obscure. Even simple formulas such as for wave speed may require more math than we want to spend class time on. DA is a good way to generate complex equations and tie them to the idea that their construction is governed by fundamental principles.

## Why aren't unit checks good enough?

While units and unit checks are often encouraged in introductory physics, it's important to stress dimensionalities as well as units. When students do unit checks, the units are associated with a number — the specific value the symbol has. If we want them to learn to "think in the blend" it helps to focus on dimensionalities rather than units. Dimensionalities tell us *what kind* of choices we get to make in assigning numbers to physical quantities; units tell us *which specific choices* we have made. There are significant cognitive differences to what a dimensional analysis activates compared to what a unit check activates.

For a given dimensionality we often work at multiple scales — a length can be meters, centimeters, microns, nanometers — keeping track of specific units adds an additional cognitive load. It's essential when you're doing a numerical calculation, but counter-productive when you're trying to think about what an equation means physically.

It's useful to focus on dimensions when helping students build cognitive blends between physics and math. It's useful to focus on units when students are learning to quantify and estimate.[10] Dimensional analysis becomes particularly valuable in situations where functional dependence and scaling are important.

### The notation for dimensionality can be a problem

Since dimensionalities specify what *kind* of thing a symbol represents and not the specific value, dimensionalities have no numbers attached to them. This makes the algebra of dimensionalities different from standard algebra.

The standard notation for representing a dimensionality of a symbol is to put the symbol in square brackets and equate that to a mix of capital letters representing the different possible measurement tools: L for length, T for time, M for mass, Q for charge, and $\Theta$ for temperature.

I like to say,

*In dimensional analysis, the square brackets ask the question to what's inside them: What kind of measurements were combined to create the number assigned to you?*

Thus, the statement that velocity is obtained by taking a length measurement and dividing it by a time measurement is $[v] = L/T$. Note that the bracket is an operator, not a marker. It goes around variables and parameters to indicate that it is yielding a dimension, not something with a number attached. See the example below.

While this is the standard notation, students have some trouble with the idea that dimensional expressions are not standard algebra. For example, in dimensional analysis the following are correct equations.

$L + L = L$   (the sum of two lengths is a length)

$2L = L$      (if you double a length it's still a length)

Until they understand what information DA is coding, students can have trouble with this idea (and may put numbers into a dimensionality).[11]

## An example:
## Finding an equation

Here's an example that shows how to do a DA and how it is different from a unit check. (I do this example even in semesters when I don't cover surface tension in class.)

We expect the height, h, that the liquid rises in the capillary tube depends on the surface tension, $\gamma$, the gravitational field, g, the mass density of the liquid, $\rho$, and the diameter of the tube, d. The surface tension parameter $\gamma$ has dimensions of force/length. From physical plausibility (What should make the height larger? Smaller?) we expect the equation should look something like

$$h = \frac{\gamma}{\rho g d^n}$$

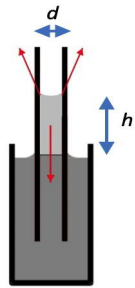

Use DA to determine the correct value of $n$.

Let's use DA to analyze the kind of measurements that go into each symbol.

- The surface tension is force/length so the dimensionality of gamma, $[\gamma]$, is that of force divided by length. Force has the same dimensionality as "ma". Putting in the dimensionality of mass (M) and acceleration ($L/T^2$)
  $[\gamma] = [F]/L = [ma]/L = (ML/T^2)/L = M/T^2$
- The density ($\rho$) is mass/volume so $[\rho]$ is mass (M) divided by that of volume (length cubed) or
  $[\rho] = M/L^3$.
- The gravitational field looks like an acceleration so
  $[g] = L/T^2$.
- The distance, d, is a length, so the $[d] = L$.

We want h to have dimensionality L, so when we multiply all the dimensionalities for the equation together we need to get L. Let's see what we get doing one kind of measurement at a time.

Only $[\gamma]$ and $[\rho]$ have a mass in them and one each. Since for $[\gamma]$ M is in the numerator and for $[\rho]$ it's in the denominator, their M's cancel:

$$[\gamma]/[\rho] = (M/T^2) / (M/L^3) = L^3/T^2.$$

Only $[\gamma]$ and $[g]$ have a time in them and each has a $1/T^2$. Since $\gamma$ is in the numerator and g is in the denominator, their T's cancel:

$$[\gamma]/[\rho g] = (L^3/T^2) / (L/T^2) = L^2$$

We only a length left in the denominator, so if we want h to be a length, we want the denominator to be a 1/L. $[\rho]$ has $1/L^3$, the $[g]$ has an L so together they are a $1/L^2$. To get a 1/L the $[d^n]$ has to cancel one (and only one) of those L's, so d has to be to the first power. Therefore n must be 1 to get $[h]$ to come out as just a length.

This result is true no matter what units you choose. Our analysis has only focused on the *kind* of physical quantity each symbol stood for, a specific application of the conceptual blend. A detailed groupwork lesson walking students through a DA is given in the supplementary materials.

## Dimensionality is a choice

In introductory physics, we're used to working with the 5 measurement dimensions listed, but we may be unsure the extent to which these are "real". Though there is strong pressure to maintain a standard (SI), many of us trained in physics have been exposed to different choices — for example, taking the speed of light equal to 1, giving space and time the same dimensionality, or choosing units so you don't have to introduce an independent dimension of charge (electrostatic unit system), or measuring temperature as energy (in electron volts) so you don't have to introduce an independent temperature dimension.[12]

While that's interesting, it's not really relevant in helping students to learn to think about physics. A better way is to consider dimensionalities as a choice that we make to help us map physical concepts into mathematical variables — a first step in building the mental blend of physics and math.

*Whenever we have a free choice in setting the scale of a measurement, we can choose to define a dimensionality. This is convenient in helping us think about the structure of equations and how they represent physical meaning.*

Dimensionalities guarantee *that our equations do not lose validity when we change our arbitrary choices.* It's really about how we are thinking about how physics relates to mathematical symbols. When we shift our focus — for example to systems where we have to take into account special relativity — we might well decide to change how we want to assign dimensionalities in that situation and treat time and space as the same.

This flexible ("It's my choice") view of dimensionalities means that while we *can* define a dimensionality every time we make a choice of scale, we don't have to. If we're never going to use a different unit, we don't have to bother with a dimension. For example, we usually don't bother to assign a dimensionality to an angle (but we probably should).

***Do angles have dimensions?*** We often say (and it is even an accepted possibility that in the SI system of units) that if we measure angles in radians, they are considered dimensionless. That's because the radian is defined as the ratio of two lengths: $\theta$ = (arc length)/(radius) = $s/r$. Since it's a ratio of the same kind of quantities, the dimensionalities cancel out. But this misses the critical issue. Conceptually, angle is about the opening between two straight lines joined at a point. There is a choice: How many divisions do we divide a right angle into? We may choose to use other non-radian units where we choose to divide the right angle into 90 parts (the degree) or 100 parts (the gradian) instead of into $\pi/2$ parts (the radian). The $\theta = \alpha s/r$ where $\alpha$ sets the unit.

To my way of thinking, this free choice of measurement scale implies that we should assign that measurement a dimensionality. The way we typically treat angles — saying they have units but not a dimensionality — can be confusing. Students often mess up calculations of angular variables. However, attaching a dimensionality to an angle is not widely done and students tend to be familiar with radians so sticking to radians (choosing $\alpha$ = 1 radian) is probably workable. My position is somewhat controversial and not in line with the AAPT recommendation.[13] This isn't a battle I intend to take on.

## Using DA in class

Dimensional analysis is often either taught explicitly at the beginning of an introductory class or discussed in the first chapter of the text that students rarely read, but students are rarely taught how to use it and are rarely asked to do it as a specific task in the class. I believe that at least in part this is because we, as physicists, have internalized the idea so well. We blend dimensionality and units effortlessly and we frequently scan our equations to check for them. We feel that once we tell our students to watch out for units, it should all be obvious!

But students have a lot of trouble with DA. It asks them to look at symbols in a way with which they have little or no experience. They're not sure that it will help them get "the answers" (which they think are numbers), so they tend to be not only unmotivated to learn it, but resistive. The only way to reset their (epistemological[14]) expectations is to make it a part of what they are required to do and on which they are evaluated. If we never explicitly ask them to do DA in a situation in which they are evaluated, it sends the message that we don't really think that it's important.

I present DA in the first few classes. To show my students that I care about DA early in the class, my weekly quizzes have a DA problem in most weeks. I also give my students DA problems at various points through the class, especially when a concept with a new dimensionality is introduced. By the end of the year, a significant fraction of my students mention DA as one of the important things they have learned and, after the class, many pre-meds have come by to say that it was of particular value in their studying for and taking the MCAT.

Here are a couple of examples. I have included the groupwork problem, a longer homework problem on DA, and an exam problem in the supplementary online materials.[15]

### Two quiz questions

> The dipole is an important source of electric effects in fluids. It has an electric effect that is like that produced by two opposite charges separated by a small distance. The measure of the strength of an electric dipole is the dipole moment, p. The magnitude of the electric force exerted by a dipole on a charge q a distance r away from it is given by (if the dipole is correctly oriented)
>
> $$F_{p \to q} = k_C qp/r^3$$
>
> What is [p] (the dimensionality of the dipole moment)? Express your answer in terms of the dimensions M, L, T, Q, $\Theta$.

> The Reynold's number for a sphere moving in a fluid is the ratio of the drag to the viscous forces the sphere feels. It is given by the equation $Re = \left(\frac{\rho}{\mu}\right)(Rv)$. In this equation, the quantities in the first parentheses are properties of the fluid (its density, $\rho$, and its viscosity, $\mu$) and the quantities in the second parentheses are properties of the object (its radius, $R$, and its speed, $v$). What is the dimensionality of the quantity pertaining to the fluid, $\left(\frac{\rho}{\mu}\right)$?

While I've focused on the importance of dimensional analysis as building a conceptual blend for introductory physics students (especially non majors), it can also be of great value to emphasize if for physics majors throughout the curriculum, even through graduate school. It can be used, for example, in generating new ideas for equations, analyzing the structure of complex equations, or for finding natural scales to set up dimensionless equations for a numerical computation.[16]

## Instructional resources

Many of the ideas for this series of paper were developed in the context of studying physics learning in a class for life-science majors. A number of problems and activities using dimensional analysis are offered in the supplementary materials to this paper. A more extensive collection of readings and activities from this project on the topic of dimensional analysis is available at the *Living Physics Portal*,[17] search "Making Meaning with Mathematics: Dimensional analysis." Acknowledgements

I would like to thank the members of the UMd PERG over the last two decades for discussion on these issues. I thank Wolfgang Losert for the suggestion that stoichiometry and DA are conceptually similar that led to the long homework problem in the supplementary materials. The work has been supported in part by a grant from the Howard Hughes Medical Institute and NSF grants 1504366 and 1624478.

---

[1] E. F. Redish, Using math in physics: Overview; preprint.

[2] In this article, I use the term *dimensionality* rather than dimension to differentiate from the number of spatial dimensions we happen to be considering, since in introductory physics, we often do problems on a line (1D), in a plane (2D), or in a volume (3D). Having even a slightly different term can help avoid confusion.

[3] For a discussion of e-games, see the overview paper or J. Tuminaro and E. F. Redish, Elements of a Cognitive Model of Physics Problem Solving: Epistemic Games, *Phys. Rev. STPER*, **3**, 020101 (2007). 22 pages.

[4] *Scientific Foundations for Future Physicians* (Assoc. of Amer. Medical Colleges, 2009), Competency E1, p. 23.

[5] The NEXUS/Physics wiki, https://www.compadre.org/nexusph/

[6] Failing to do so can produce expensive tragedies. http://www.cnn.com/TECH/space/9909/30/mars.metric.02/

[7] Some of you will recognize this as group theory. This is an example of how we use a physics/math conceptual blend to replace some very sophisticated mathematical ideas. See M. Hammermesh, *Group theory and its application to physical problems* (Dover, 1989).

[8] J. Kotre, *White Gloves* (Norton, 1996).

[9] E. F. Redish, Using math in physics: 5 - Functional dependence and scaling; preprint

[10] E. F. Redish, Using math in physics: 2 - Estimation; preprint.

[11] I'd love to use small measurement icons instead of "L, M, T, Q, Θ" - something like shown in the figure above. But despite the popularity of emoticons with our students, this is not likely to catch on.

[12] There is a whole field of metrology that is concerned with the issue of how many independent measurements we need and what's the best way to define them.

[13] The radian — That troublesome unit, The AAPT Metric Education and SI Practices Committee, G. Aubrecht et al., The Physics Teacher 31, 84 (1993); https://doi.org/10.1119/1.2343667; see also Dimensionless units in the SI, P. Mohr & W. Phillips, *Metrologia* 52:1 (2015) 40-47.

[14] By "epistemological" I mean "knowledge about knowledge." Students' misconceptions about the kind of knowledge they are learning and what knowledge they need to bring to bear in the class are often responsible for student difficulties and resistance. See ref. 2 for a discussion.

[15] EPAPS link

[16] Some readings and problems for a junior level class in *Methods of Mathematical Physics* are available at http://umdperg.pbworks.com/w/page/34231836/Methods-of-Mathematical-Physics, choose Content / Things of Physics / Dimensions.

[17] *The Living Physics Portal*, https://www.livingphysicsportal.org/